# Effect of finite terms on the truncation error of Mie series

Antonio Alvaro Ranha Neves,[1,*] and Dario Pisignano[1,2]

[1] *National Nanotechnology Laboratory (NNL) of CNR, Nanoscience Institute c/o Distretto Tecnologico, Università del Salento, Via Arnesano 16, Lecce 73100, Italy*
[2] *Dipartimento di Ingegneria dell'Innovazione – Università del Salento, via per Arnesano, 73100 Lecce, Italy and Center for Biomolecular Nanotechnologies, Istituto Italiano di Tecnologia @UniLe, via Barsanti, 73010 Arnesano, Lecce, Italy*
*Corresponding author: aneves@gmail.com

The finite sum of the squares of the Mie coefficients is very useful for addressing problems of classical light scattering. An approximate formula available in the literature, and still in use today, has been developed to determine a priori the number of the most significant terms needed to evaluate the scattering cross section. Here we obtain an improved formula, which includes the number of terms needed for determining the scattering cross section within a prescribed relative error. This is accomplished using extended precision computation, for a wide range of commonly used size parameters and indexes of refraction. The revised formula for the finite number of terms can be a promising and valuable approach for efficient modeling light scattering phenomena. © 2012 Optical Society of America

Light scattering today is still an important and active field, whose application has largely outgrown the field of atmospheric optics, affecting several issues in optical trapping [1,2], metamaterials and cloaking [3], plasmonics [4], and realistic physical based rendering [5]. In particular, the phenomenon of plane waves scattering by a spherical body is exactly described by Mie theory in terms of an infinite series [6]. For the purpose of numerical computation, only a finite number of terms are retained, based on the size of the scatterer in relation to the wavelength. The number of terms conserved is commonly determined by Wiscombe's criterion [7]. This method has been initially proposed to compute the spherical functions by downward recursion, and to utilize the new vector processing technology of the time, which required a number of terms established *a priori*. Nowadays, with the increased computational power and capability of data storage, this is not an issue, hence the results from Mie simulations could be more effective, reaching an optimal compromise between computational time and precision. In this letter, we revisit the calculations and provide an adequate criterion for terminating the infinite sum in terms of the required precision. These approximations can offer valuable approaches to model light scattering phenomena.

Besides the truncation error, in numerical computation one has to take into account also the round-off error, due to the finite precision of machine numbers as in floating-point representation in use by computers. To avoid round-off errors, the method herein proposes to use a computational tool (Mathematica 8, Wolfram Research Inc.) that allows an extended-precision computation to be evaluated. Working with a precision, that is orders greater than the truncation error, the round-off errors due to finite precision computation becomes negligible to our problem. Moreover, the needed spherical functions (including the functions having arbitrary complex argument and high order, say ≥ 100 or more depending on the scatterer size) can be readily obtained at the desired precision, and tracked during the computation.

The theory that treats the scattering of a linearly polarized monochromatic plane electromagnetic wave by an arbitrary uniform sphere in a non-absorbing medium, known as the Mie-scattering formalism, is a rigorous solution of the Maxwell equations and contains all the effects that contribute to the scattering [8,9]. According to this theory, the normalized scattering efficiency is related to the partial-wave amplitudes by,

$$Q_s = 2x^{-2} \sum_{n=1}^{\infty} (2n+1)\left(|a_n|^2 + |b_n|^2\right), \qquad (1)$$

where $a_n$ and $b_n$ are the Mie scattering coefficients. In the case of a homogeneous sphere, written in term of the logarithmic derivative the latter coefficients are [10],

$$a_n = \left\{\frac{j_n(x)}{h_n^{(1)}(x)}\right\} \frac{\ln'\left[mxj_n(mx)\right] - m\ln'\left[xj_n(x)\right]}{\ln'\left[mxj_n(mx)\right] - m\ln'\left[xh_n^{(1)}(x)\right]}, \quad (2)$$

$$b_n = \left\{\frac{j_n(x)}{h_n^{(1)}(x)}\right\} \frac{\ln'\left[mxj_n(mx)\right] - m^{-1}\ln'\left[xj_n(x)\right]}{\ln'\left[mxj_n(mx)\right] - m^{-1}\ln'\left[xh_n^{(1)}(x)\right]}. \quad (3)$$

In the previous expressions, the derivative is with respect to the argument, $j_n(x)$ are the spherical Bessel and $h_n^{(1)}(x)$ are the spherical Hankel function of the first kind. The ratio of the refractive index of the particle (which could be complex) to the medium is $m$, $x$ is the size parameter defined as $2\pi a/\lambda$ (where $a$ is the sphere radius and $\lambda$ is the wavelength in the medium outside the particle). One should notice that the only difference between Eq. (2) and Eq. (3) is the dependence on the index of refraction term ($m$) in the last term of the numerator and denominator.

The problem of applying the expression reported in Eq. (1) directly in a computational scheme is the need for an infinite number of terms. In fact, the truncation of the multipole expansion to a finite number of term, $N$, is generally adopted, which depends on the convergence of the Mie coefficients through the spherical functions. The terms responsible for the convergence of the Mie coefficients are emphasized, in Eq. (2) and Eq. (3) with the curly brackets, whose arguments are always real. As $n \rightarrow \infty$, the modulus of the Mie coefficients tends to zero, due mainly to the terms in the curly brackets [11]. The relative error of the scattering efficiency due to truncating the infinite series, is:

$$\Delta Q_s / Q_s = 1 - 2x^{-2} [Q_s]^{-1} \sum_{n=1}^{N} (2n+1)\left(|a_n|^2 + |b_n|^2\right). \quad (4)$$

The stopping criterion for the infinite sum (Eq. (1)), according to Wiscombe criterion, was chosen to be $|a_n|^2 + |b_n|^2 < 5 \times 10^{-14}$, due to previous reported work [12]. Instead, here we adopt the relative error, which is normalized (Eq. (4)). As known, the Wiscombe's criterion is an empirical fitting yielding a stepwise function, but in its absolute form, is the integer part of,

$$N = x + 4.05 x^{1/3} + 2. \quad (5)$$

In Eq. (5), the first term on the right hand side ($x$) is related to the localization principle [13], and mathematically, it represents the region where the spherical function contributes the most. The second term ($x^{1/3}$) corresponds to the contributions of the surface waves [14]. An equation with the same functional form as Eq. (5), but which depends on the truncation error is desired.

The Mie coefficients depend on the spherical functions, and these can be approximated in different domains. In particular, many authors apply asymptotic formulae for values of $x$ such that $|x| \gg n$. In our case, the domain of interest is right outside the spherical body, as suggested by Eq. (5), for $|x| \cong n$, where the asymptotic formula converges slowly. In this transition region, a suitable approximation for the spherical functions can be expressed as asymptotic series and in terms of Airy functions [14,15,16], as:

$$j_n(x) \approx (2x)^{-1/2} v^{-1/3} (8c)^{-1/4} Exp\left(-2\sqrt{2} c^{3/2}/3\right), \quad (6)$$

$$y_n(x) \approx -(x)^{-1/2} v^{-1/3} (8c)^{-1/4} Exp\left(2\sqrt{2} c^{3/2}/3\right), \quad (7)$$

where $v = n + 1/2$ and $x = v - cv^{1/3}$. Combining these approximations for the Mie coefficients in Eq. (4), and using $\Delta Q_s / Q_s = 10^{-\varepsilon}$, after some algebraic manipulation, taking into account the dominant term, the following functional dependence is obtained:

$$N \approx x + \alpha \varepsilon^{2/3} x^{1/3} + \beta. \quad (8)$$

In the previous equation, $\alpha$ and $\beta$ are numeric constants to be determined by fitting Eq. (8) to simulated runs of typical data. The truncating error of the scattering efficiency (Eq. (4)) is then generated as a function of the number of terms in the series, for a wide range of commonly used material refractive index, $1.1 \leq \text{Re}(m) \leq 2.0$. The chosen points for the size factor, within the range $1 \leq x \leq 1000$, corresponds to the real part of the poles for the Mie scattering coefficient, thereby taking into account the spikes (ripple structure) in the scattering cross section due to the morphology dependent resonances (MDR). The center of a MDR (or whispering gallery modes) is conveniently determined from the position (size factor) where $\text{Im}(a_n)$ (or $\text{Im}(b_n)$) changes sign [17].

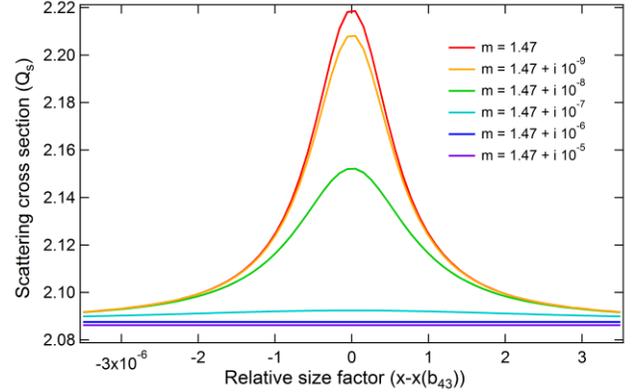

Fig. 1. (Color online) Scattering cross section for different values of imaginary refractive index, for the $b_{34}$ resonance, centered near $x = 40.32638$ as in Ref. 17.

The known effects on the MDRs, for the case of increasing imaginary part of the refractive index (small absorption), is a rapid decrease in the height of the peak of the scattering cross section, with a concomitant increase in the resonance width. As a result, large imaginary refractive index has negligible contribution to the needed amount of term in the series, and the maximal value for the scattering cross section is given by the pure dielectric microsphere, evidenced for resonance $b_{34}$ (i.e. $n = 34$) in Fig. 1. As for the role of the convergence, it shows little dependence on the dielectric refractive index, requiring a maximum of 2 extra terms when switched from $m = 2.0$ to $m = 1.1$ in the investigated range. This is expected, since the convergence term, in Eqs. (2-3), does not depend on the relative refractive index, and this is why it is not included in the empirical Eq. (5).

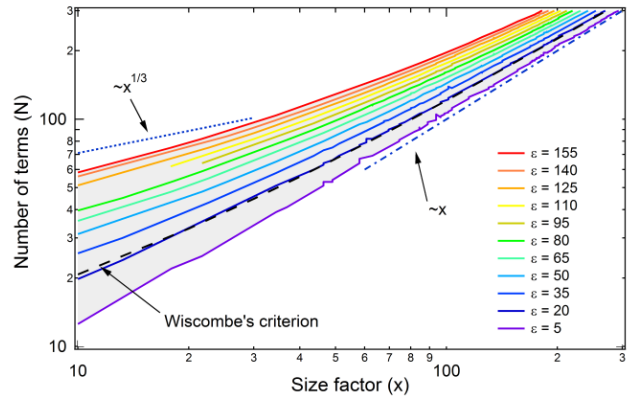

Fig. 2. (Color online) Log-Log plot evidencing the functional dependence of the maximum number of terms ($N$) as a function of size factor ($x$). Each curve is plotted for different truncation errors from $10^{-5}$ to $10^{-155}$, from bottom to top respectively. Dashed and dot-dash blue lines represent trends while the dashed black line represents Wiscombe's criterion Eq. (5).

The main result of our simulations are presented in Figs. 2 and 3, which show the number of required terms as a function of size factor for different values of the truncation errors. The fitting of Eq. (8) would be done for the maximal number of terms in the investigated range, i.e. for $m = 2$ and where the size factors are MDR points.

Firstly, an important finding consists in the fact that the number of terms for smaller size factor deviates from the linear dependence, as the truncation error increases (higher curves). Actually for smaller size factor and smaller truncation error, the term proportional to the third power in Eq. (8) becomes more significant. As for the rate of increase in the number of terms considered in the series, with the reduction of the truncation error, it scales as $\varepsilon^{2/3}$.

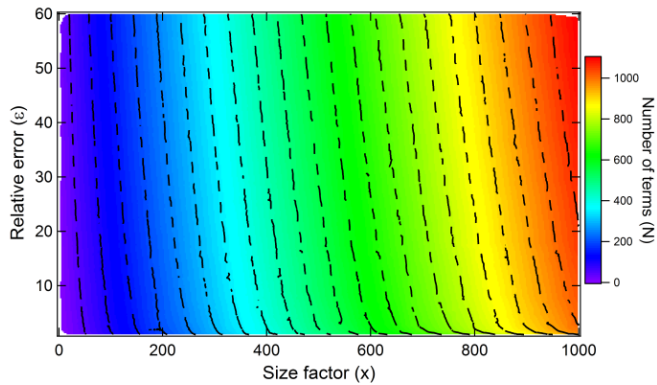

Fig. 3. (Color online) Contour plot of the number of terms ($N$) as a function of size factor ($x$) and relative error ($10^{-\varepsilon}$). First contour (left) corresponds to ($N = 50$), each successive contour adds another 50 terms to the series.

Fitting of Eq. (8) to several generated data points (Fig. 3) was done within a smaller range of investigated parameters, $1 \le x \le 200$ and $\varepsilon < 50$, a condition satisfied in most of the current applications, yielding:

$$N = x + 0.76\varepsilon^{2/3} x^{1/3} - 4.1. \qquad (9)$$

Incidentally, we notice that a direct comparison of the expressions above with the original criterion, Eq. (5), is hard to be performed, since it depends on the size parameter, but within the examined range it would correspond for $\varepsilon \approx 15$ (dashed line in Fig. 2).

In summary, the work presented in this letter has investigated the number of terms needed as a function of the truncating error to determine the scattering efficiency of a homogeneous spherical scatterer by a plane wave. From the practical point of view, the proposed expression, (Eq. (9)), is clearly useful to facilitate fast simulations of light scattering phenomena within a predetermined maximal error. It should be noted that this termination criterion applies to homogeneous spheres only, whereas other shapes result in their own scattering coefficients and therefore in specific convergence criteria. For instance, these could be exploitable in various light scattering phenomena include optical trapping of free-standing nanostructures, such as polymer nanofibers and other nanoparticles [18,19]. We anticipate that the here suggested, improved terminating approach will be helpful for researchers developing efficient light scattering code, balancing precision with computational time.

This research was partially supported by the Network of Public Research Laboratories #13 (MITT) of Apulia Region and by the FIRB Projects RBNE08BNL7 "Merit" and RBFR08DJZI "Futuro in Ricerca".